\documentstyle[12pt]{article}
\begin{document}
\begin{center}
 \huge Chaos in Periodically Perturbed 
Monopole $+$ Quadrupole Like Potentials.
\end{center}

\vspace{3ex} 

\centerline{  \em P. S. Letelier\footnote{e-mail: letelier@ime.unicamp.br}
 and 
W. M. Vieira\footnote{e-mail: vieira@ime.unicamp.br} } 
\vspace{1ex}

\begin{center}
 Departamento de Matem\'atica Aplicada-IMECC\\
Universidade Estadual de Campinas\\
13081-970 Campinas,  S.P., Brazil \\ \vspace*{0.3cm} \end{center}

\baselineskip 0.7cm
 \vspace*{6ex}
  The motion of a particle  that suffers the influence of simple
 inner (outer)  periodic 
perturbations when it evolves  around 
a center of attraction  modeled by an inverse square law plus
a quadrupole-like term  is studied.  
The   equations of motion  are used to reduce
the Melnikov method to the study of simple graphics. 

 \noindent
PACS numbers: 0.5.45.+b, 03.20.+i, 95.10. Ce.
 
\newpage
\baselineskip 0.6cm
\noindent
{\bf 1. Introduction.}

 \,\,Since the pioneering work of
 Poincar\'e \cite{poincare} in celestial
mechanics wherein the mathematical basis of  deterministic chaos
 was  established, the study of  homoclinic phenomena has been the 
focus of increasing attention
not only in celestial mechanics but in many different branches of
 physics, chemistry and biology \cite{bailin}. 
 Due to its universality and 
 intrinsic mathematical interest, models in which unstable
periodic  orbits (UPOs) are subjected to  small 
periodic perturbations   has been one
 of the main paradigms of deterministic chaos \cite{arnold}. An 
  analytical tool
to study these models is the Melnikov function that describes the 
transversal distance between the unstable and stable manifolds 
emanating from an UPO. Its isolated odd zeros indicate the crossing
 of these manifolds, hence the onset of chaos \cite{melnikov}. The use
 of Melnikov method in connection to 
Kolmogorov-Arnold-Moser (KAM) theory for a large class
 of perturbations was studied by Holmes and  Marsden \cite{HM82}, see also
\cite{holmes}.

Examples of applications of the Melnikov method in gravitation
 are: the motion of a nearly symmetric heavy top \cite{HM83},
the motion of particles in a two-dimensional St\"ackel potential
 separable  with $\cos{m\varphi}$
 perturbations ($m=1,2,...$)
and other perturbations \cite{gerhard85}, extensions of the former work for 
perturbation of the three-dimensional St\"ackel potentials \cite{gerhard86}   
the gravitational collapse
 of cosmological  models \cite{ivano}, the
 study of orbits around a black hole perturbed by either gravitational
 radiation \cite{BC,LV97} or an external quadrupolar shell \cite{moeckel}

In this paper we  consider the equatorial motion of a particle moving 
in a potential   modeled by an inverse square law plus
a quadrupole-like term.  This potential describes more 
realistically the gravitational force of  attraction
of stars (like the sun) on  planets  and 
galaxy bulges
that to a certain degree of approximation can be represented
 by a monopole plus a quadrupole on stars outside the galaxy.
 Also, this  potential 
  arises naturally in  General Relativity 
in the study of test particles in a spherically symmetric
 attraction center (Schwarzschild black hole), the ``quadrupole" 
being associated to the test   particle angular momentum, see
 for instance \cite{straumann}. 

 Next, we consider a class of periodic perturbations of this
 integrable planar motion. The perturbations
  can be the result of periodic motion of
 mass distributions either inside or outside the orbit of interest.
 The fact that we were able to integrate in a closed form the 
 equation of motion of  the UPO is used to 
reduce the analysis of the Melnikov function to the study 
of simple   graphics. We 
find that the perturbations will originate chaos in a 
 variety of situations 
 and range of parameters. Finally,  we argue that chaotic behavior 
 is generic for such a perturbations.

\vspace{0.5 cm}
\noindent
 {\bf 2. The homoclinic orbit.} 

We shall consider the orbit of a 
particle of mass $m$ moving on a plane under the influence of a
 force described by a potential with a monopolar term plus a 
quadrupole-like contribution not 
necessarily small,
\begin{equation}
V=-km/R-Qm/R^3 , \label{V}
\end{equation}
where $k=GM$, 
$M$ is the mass of the attraction center, and $Q$ is the 
quadrupolar strength. In astronomy, 
$Q=GMJ_2 R_{0}^{2}/2$, with   $R_0$ being a  linear  measure of the size
 of the central body and  $J_2$ the distortion
 parameter, see for instance \cite{straumann}.  All 
the undefined symbols have their usual meaning. Note that for a body with
reflection symmetry the octupolar term is null ($J_3=0$).

In General 
Relativity, $Q$ is proportional to the
 square of the angular momentum  of the test particle \cite{straumann} .
 
The Lagrangian  of the particle is  
\begin{equation}
  L=\frac{m}{2} \biggr[ \biggr( \frac{dR}{dt} \biggl)^2+
R^2\biggr(\frac{d\phi}{dt}\biggl)^2+\frac{2k}{R}
+\frac{2Q}{R^3} \biggl] \label{lag}.
\end{equation}
Since $\phi$ is a cyclic variable we have the conservation
 law $h=R^2 d\phi/dt$. Hence, the motion of the particle is
 determined by the effective Hamiltonian
\begin{eqnarray}
& & H_0=\frac{P^2}{2m}+V_{eff}, \;\;\;\ (P=mdR/dt), \label{H0} \\
& & V_{eff}=m\biggl(\frac{h^2}{2R^2}-\frac{k}{R}-
\frac{Q}{R^3}\biggr) .  \label{Vef}
\end{eqnarray}
The stationary values of $V_{eff}$ are
 $ R=\frac{h^2}{2k}(1\pm \sqrt{1 -12\beta})$ with 
$ \beta=kQ/h^4$. The minus (plus) sign correspond to the relative 
maximum  (minimum). Note that
$\beta$ is a non dimensional parameter. We shall denote by
 $R_{un}$ the UPO position  given by the stationary value with minus 
sign (the maximum); we find that
$
V_{eff}(R_{un})=\frac{mh^2}{6R_{un}^2} (-1+2\sqrt{1
 -\beta}).$
Since, for large $R$, the potential $V_{eff}\sim -\frac{k}{R} \leq 0$, we 
need $V_{eff}(R_{un}) \leq 0$  to 
have a periodic  motion of the particle limited by
 $R_{un}$ and  $R_m$,  the last being the turning point.  Hence the
 parameter $\beta$ is 
limited by $1/16=0.0625\leq \beta \leq 1/12\sim 0.0833.$ 

It is  convenient to work with dimensionless quantities. 
 We shall redefine our variables using the constants   of the
 problem in the following way:
$r=kR/h^2 , \,\, \tau=k^2 t/h^3  ,   \,\, 
 \varepsilon= h^2 E/mk^2,$
where  $E=H_0$ is the total energy of the particle. We recall
 that these quantities are proportional, respectively, to the
 radius, the period, and the energy of a particle of mass $m$ 
moving in a circular orbit with areal velocity $h$ around
 an attraction center
 characterized by $k$.  The 
energy  integral can be written as
\begin{eqnarray}
& & 2 \varepsilon =\dot{r}^2+ 2 v_{eff}, \label{E2}\\
& & 2 v_{eff}= 1/r^2-2/r-2\beta/r^3,  \label{vef}
\end{eqnarray} 
where the overdot indicates derivation with respect to
 the time parameter $\tau$. 
In these new units, $R_{un}$  is written as  
 $r_{un}=(1-\sqrt{1 -12\beta})/2$. For $\beta$ in the
 interval $1/16\leq \beta \leq 1/12$
  we have $1/4 \leq r_{un}\leq 1/2$. In
 Fig. 1 we present  a graphic of the potential
(\ref{vef}) for the values of $\beta=$ 0.068 (top curve), 0.072,
 and 0.078 (bottom curve), the maximum value is located at $r_{un}$. 
A particle with energy  $\epsilon =v_{eff}(r_{un})$ will describe 
either the UPO itself ($r=r_{un}$, $\dot{r} =0$) or the homoclinic orbit
tending to the UPO as $\tau \rightarrow \pm \infty$. This
 homoclinic  orbit is enclosed
by  circles of radii $r_{un}$ and
 $r_m=2r_{un}(1-r_{un})/(4r_{un}-1)$. Now, by making
  $\epsilon =v_{eff}(r_{un})$ we can write (\ref{E2}) in the 
equivalent form
\begin{equation}
\frac{r^{3/2} \dot{r} }{(r-r_{un})\sqrt{r_m -r} }= \pm w_\beta ,
  \label{E3}
\end{equation}
with 
\begin{equation}
 w_\beta=\sqrt{(4r_{un} -1)/(3r_{un}^{2} )} \label{w}.
\end{equation}
  The differential equation  (\ref{E3}) admits the  quadrature
\begin{eqnarray}
\mp w_{\beta}\tau=\sqrt{r(r_m-r)}+(r_m+
2r_{un})\arctan\sqrt{\frac{r_m-r}{r}} \nonumber& &\\
+ \frac{ 2r_m^{3/2} }{ \sqrt{r_m- 
r_{un}} } {\rm arctanh} \sqrt{ \frac{ (r_m-r)r_{un} }
{ (r_m-r_{un})r } }. \label{tau}&  &
\end{eqnarray}
We have chosen the constant of integration to have 
$\tau=0$ at $r=r_m$.
A graphic $\tau(r)$ of  the positive branch of (\ref{tau}) 
is pictured in Fig. 2 for  values of the
 parameter $\beta=$ 0.064 (top curve), 0.065, 0.066, 0.067,
 and 0.068 (bottom curve). We 
see that the particle takes a finite time to travel
 from $r_m$ to the vicinity of
 $r_{un}$ and then  an infinite time to arrive (depart) to (from) 
$r_{un}$ itself, wherein is located the UPO.

\vspace{0.5 cm}
 \noindent
{\bf 3. The Melnikov method.} 

Let us consider an integrable  Hamiltonian 
 $H_0=\frac{p^2}{2m}+ V(q), $
 which  admits at least one UPO with the corresponding homoclinic  
 orbit,  and also a small periodic
 perturbation described  by the Hamiltonian function
 $\eta H_1(q,p,t)$. Then the transverse distance in phase
 space between the unstable  and the stable manifolds emanating
 from the UPO is proportional to
\begin{equation}
M(t_0)=\int_{ -\infty}^{ \infty} {\{ H_0,H_1\}dt}, \label{M}
\end{equation}
where the integral is taken along the unperturbed homoclinic orbit, $t_0$ is
 an arbitrary initial time, and $\{f,g\}$ are the usual Poisson
brackets, see for instance \cite{holmes}. If  there is an
 intersection for some $t_0$, i.e., an isolated odd  zero of
 $M(t_0)$, then there will be one for every  $t_0$. This infinitely
 many crossings of manifolds   will produce a tangle that is the signature 
of homoclinic chaos \cite{melnikov,holmes}. 

The unperturbed system is
  Eqs. (\ref{E2}) and (\ref{vef}) with $H_0=\varepsilon$, 
 and $(q,p)=(r,\dot{r})$. For our 
purposes, it is enough to consider  periodic  perturbations of the 
particular form  $H_1=f(r)\cos(\Omega\tau)$.
 Perturbations of this form are  representative of a 
very general situation: take  the potential of a mass distribution 
moving periodically and make a Fourier expansion in time and a multipolar
 expansion in the space variables. Next, consider approximate axial
   and  reflection symmetry. Therefore, the 
generic term of the series expansion in the plane of the
 UPO ($\theta=\pi/2$) will look like $H_1$ above with $f(r)$ a 
positive (negative) powers of $r$ for mass distributions
 outside (inside) the homoclinic orbit.
   We find for 
the Melnikov function (\ref{M}),
\begin{equation}
M(\tau_0) = \int_{ -\infty}^{ \infty} {\frac {d r}{d \tau} 
\frac{ df(r)}{dr} \cos[\Omega(\tau-\tau_0)] d\tau} .\label{M2}
\end{equation}

In our case, for the homoclinic orbit (\ref{tau}), we get
\begin{eqnarray}
& & M(\tau_0) = -2K(\Omega)\sin(\Omega\tau_0)   ,\label{M3}\\
& & K(\Omega)=\int_{r_{un}}^{ r_m} {\frac{ df(r)}{dr} 
\sin [\Omega\tau(r)] dr}, \label{I}
\end{eqnarray}
where in the integrand $\tau(r)$ means the positive 
branch of (\ref{tau}). Note that in (\ref{M3}) does not appear 
a term proportional to $\cos(\Omega \tau_0)$, its coefficient 
  being null due to the fact that 
the homoclinic  orbit has reflection symmetry with respect to
 the  $r$-axis, and $\cos(\Omega\tau)$ is an even function 
in the  $\tau$ variable.
Thus the Melnikov function will have
 the required zeros as long as $K(\Omega)\not= 0$.
The coordinate change $\tau\rightarrow r$, instead of
the usual  $\tau\rightarrow \phi$, allows 
 us to pass from an infinite interval to a finite one in (\ref{I}).   
This observation will play a key role in the evaluation of the function
$K(\Omega)$.

\vspace{0.5 cm} 
\noindent
{\bf 4. Particular cases.}

Firstly, we shall consider the
 function $K(\Omega)$ for exterior  perturbations of the type
 $f(r)=r^n$ with $n=1,2,... $. For a particle moving
 around the attraction center, 
they model  periodic perturbations due to  mass
distributions  beyond the orbit of the particle. $n=1$ 
represents dipolar contributions, $n=2$  quadrupolar, etc. Static
 quadrupolar and octopolar exterior  contributions, inspired in the well known  
 H\'enon-Heiles model, give rise also to chaotic behavior in  
 General Relativity \cite{VL96}.

Now, we shall study the integrand of  $K(\Omega)$ formed by the product of an 
 oscillating function by a polynomial.  Near
 $r =r_{un}$ this  oscillation is a rapid one. To better understand this 
behavior  we display 
in Fig. 3  a graphic of $\sin [\Omega\tau(r)] $
 for $\beta=0.064$ and $\Omega$ =0.05 (top curve), 0.06, 0.07 and 0.08
 (bottom curve). For $\Omega>0.08$ we will 
have more zeros in the interval $1<r<10$ than  for $\Omega$ =0.08. For
 $\Omega<0.05$ the curves
 will look like the one for $\Omega=0.05$. Then it is easy to see that the
 integral $K(\Omega)$ 
for $f(r)=r$ will  be non zero for $\beta=0.064$ and $\Omega<0.06$.  For
 $n>1$ we will have a more favorable situation. In Fig. 4 we
 plot the integrand of $K(\Omega)$  for $\beta$=0.064, $\Omega$=0.06, and 
  $f'(r) \equiv df/dr=r$ (top curve), $r^2/10$, and $r^3/100$ (bottom curve),
 i.e., for a quadrupolar, octopolar and $16^{th}$-polar external
 periodic perturbations, respectively. In all 
cases the area under the curve is clearly not null, then we will have 
 chaotic 
motion  in these situations.  The values of $\Omega$=  0.06 and 
0.07 were obtained from
$
\Omega_k \simeq \pi/\tau_k ,
$
where $\tau_k$  corresponds to the maximum of the 
 curvature ($k=\tau''/(1+(\tau')^2)^{3/2})$  of 
$\tau(r)$ \,\, 
 for the first value, and to 
 the inflection point for the second one. From the previous analysis
 we see that the first value of $\Omega$, the one based on  
 curvature, is better.  Thus, we   conclude that a perturbation 
with 
$f(r)\sim r^\alpha$ with $\alpha \geq 1$ and $\Omega \leq \Omega_k$
 will always be chaotic. A closer look shows that this
 criteria is good whenever $r_m -r_{un}>1$, otherwise one needs
 $\Omega \ll \Omega_k$. It is also interesting to compare the
 perturbation frequency $\Omega$ with
$\omega_{un}=1/r_{un}^2, $
that is the angular frequency ($\dot{\phi})$ of a particle at the UPO 
in dimensionless units. For $\beta=0.064$ we have $\omega_{un}$=15.
 So, $\Omega\ll \omega_{un}$.

Now we shall study   perturbations on the motion of
 the particle that 
can be modeled by functions $f(r)\sim r^{-n}$ with $n=$2, 3,...; $n$=2 
represents monopolar contributions, $n$=3 dipolar, etc. 
These type of perturbations can model forces  due to
distribution of masses with periodic motions that are placed 
inside the  motion. Also, the case $f(r)\sim \log r$,   which can
 be used to model the contribution of bodies  with a spindly form,
 will be  considered.

As in the previous case, we shall begin by analyzing a particular
 situation. In Fig. 5 we  plot the integrand of $K(\Omega)$  for 
$\beta$=0.068, $\Omega$=0.15, and $f'(r) =1/(3r^2)$ (top curve),
 $1/(9r^3)$, and $1/(27r^4),$ (bottom curve). In this case we have a rapid
 oscillation (not  shown in  Fig. 5) near the point $r_{un}$ (=0.285524)
that produces very fine spines of sizes between   [-4.0,  4.0] (top 
curve), [-4.5, 4.5],  and
 [-5.5, 5.5] (bottom curve). The areas under the positive
 parts of the curves are clearly greater than the areas under the
 negative ones, which
 is  under very fine spines. So, also in
 this case  $K(\Omega)$  will be different from zero and the Melnikov 
method assures the occurrence of chaos. The
 value  $\Omega$=0.15 was obtained from
$
\Omega_k \simeq \frac{\pi}{2\tau_k }. 
$
 It is also instructive to 
compare the value of $\Omega=0.15$ with the UPO frequency. For $\beta=0.068$ we 
 find  $\omega_{un}=12$. So, again  $\Omega\ll \omega_{un}$.
For powers $n \gg 4$ we have that the relative size of  the positive
 part of the area under the curve of the integrand of 
 $K(\Omega)$ begins  to decrease and the 
area of the spines to increase. Hence, the graphic method is
 not appropriated in this case, though one can always  numerically evaluate
 $K(\Omega)$. We will be back to this point
 later.  A similar analysis shows that for  $f(r)\sim \log r$  we also 
have a strictly positive $K(\Omega)$.
Therefore, we can say that for perturbations characterized by 
 functions $f'(r)\sim r^{-\alpha}$ with $1<\alpha < 4 $ and
  $\Omega<\Omega_k$, we will have chaotic orbits. If one further restricts 
the range of $\beta$ by the relation
$r_{m}-r_{un} < 2$, one can have chaos with
 perturbations for  any value of  $n>1$ in $f(r)\sim r^{-n}$.

\vspace{0.5 cm}
\noindent
{\bf 5. Discussion.} 

Let us first examine the domain of applicability of our results.
 The method is based in the existence of the homoclinic orbit
that restrict the parameter $\beta$ in a considerable way
($1/16\leq\beta\leq1/12$).   With the mass of the attraction center
 $k=GM$ and the areal velocity of the test mass we construct  our length 
scale $k/h^2$, which is  kept arbitrary in our analysis.  

We shall examine as a limit case the value of the parameters 
for particles moving near
the minimum of the effective potential $v_{eff}$ (cf. Fig. 1). 
In principle, these particles
 will feel to some extent  the presence of the saddle point
 (in phase space), the maximum of the potential, that is the 
responsible for the instabilities
that give rise to the chaotic motion. In particular, for a circular 
orbit   with radius $r_P=R_P h^2/k $ at the minimum we get, 
\begin{equation}
\beta=\frac{J_2 R^2_0 R^2_P}{2(R_P^2+\frac{3}{2}J_2R^2_0)^2} . \label{beta}
\end{equation} 
 We have in non dimensional units that 
$r_P=(1+\sqrt{1-12\beta})/2$. For the characacteristic length of the
central body we have $r_0=R_0 h^2/k = 2\beta/J_2$.  Since our potential 
(\ref{V}) decays with the distance we need that $r_P>r_0$.
 So the parameter $J_2$ that describes the oblateness of the central 
body plays a fundamental role in our analysis . From the limit condition 
$R_0=R_P$,  we get that $\beta=2J_2/(2+3J_2)^2$.    We recall 
that $J_2=0$ for a spherical distribution of matter, $1/4$ for a 
thin disk, $1/2$ for a ring. To be more specific let us consider
  that all the matter is concentrated in a thin disk of radius $R_0$. We 
have in this case that $0.0625<\beta<0.0661$. It is instructive to make
 a table of the values of $r_0$
and $r_P$ and $\varepsilon(r_0)$ and $\varepsilon(r_P)$ for
 different values of $\beta$.

\vspace{0.5cm}
Table 1.  Different parameters for $J_2=1/4$ a disk like configuration.
\begin{center}
\begin{tabular}{|c|c|c|c|c|} \hline\hline
$\beta$ & $r_0(\beta)$ & $r_P(\beta)$ & $-\varepsilon(r_0)$& 
$-\varepsilon(r_P)$\\ \hline\hline
0.0626& 0.708 & 0 .749  & 0.593& 1.143  \\ \hline
0.0630& 0.710 & 0 .747  & 0.594& 1.142  \\ \hline
0.0640& 0.716 & 0 .740  & 0.596& 1.141  \\ \hline
0.0650& 0.721 & 0 .735  & 0.599&  1.137 \\ \hline
0.0660& 0.727 & 0 .728  & 0.601& 1.136   \\ \hline
\end{tabular}
\end{center}
\vspace{0.5cm}
 Therefore we have some interval of allowed  energies and we can have 
 chaos   without a very fine
tuning of the initial conditions. From de values  in Table 1 we conclude
 that the potential well is rather deep. Furthermore the parameters also 
tells us that we can take for $J_2$ a value    smaller but close  to
 $1/4$. In this case we can add some structure to the disk: some
 thickness and/or a small central bulge.

In this  work we have presented a graphic method of implementing
the classic  Melnikov method  for significant classes of time
 periodic perturbations of a
 planar motion  around an attractive
 center modeled by the potential (\ref{V}).
 The
 method, a semi--analytical one, is based on the fact 
that we were able to obtain in a closed form the  equation motion of
 the homoclinic orbit and
  perform a change of variable in the Melnikov function that maps the
 infinite integration interval into a finite one.
This allows us to make predictions  about the zeros of the 
Melnikov function --- therefore about  the onset of chaos ---  for
 significant physically motivated  families of periodic perturbations.

Of course,  one can always numerically compute  the
 equivalent function $K(\Omega)$ for
any given function $f(r)$ and any value of $\Omega$ and
 $\beta$ in the range $1/16\leq \beta \leq 1/12$. Since  $\tau(r)$ is known 
explicitly,  this 
computation can easily be made  with a great precision. 
  For the perturbations here considered, $f(r)$ a 
power of $r$, and a fixed $\Omega>\Omega_k$, we have that
 $K(\Omega)$ takes   positive and negative values depending on the  
values of $\beta$.  Then  for 
certain specific values of $\beta$ we will have  $K(\Omega)=0$ and 
the Melnikov method does not apply in these cases.
Therefore it does look like that the generic
situation is chaotic, i.e.,  given a function $f(r)\sim r^\alpha$ and
a value of $\beta$,  only for a numerable set of
frequencies $\Omega$  we should  not have chaos.
To be more precise, the Melnikov method fails to 
predict chaos only for a numerable set of  perturbation frequencies $\Omega$.
  We note that seldom one can  find so easily  the
 parameters  range  of a given problem
that will produce a chaotic situation.  Therefore, based in our
 previous analysis,  we can conjecture that chaos is generic for the 
class of perturbed systems studied in this communication.  

Finally, we want to comment on the fully numerical approaches to the
 Melnikov method. In order  to prove numerically  the existence of 
crossings of the homoclinic orbits  the  analyticity of the Hamiltonian 
is required \cite{gerhard85,CR}. But numerics can give only indications
of the existence of  derivatives to a finite order. Then the use of
 the Melnikov method 
in this case is not in conclusive. 

 The authors thank CNPq and FAPESP for financial support.

\newpage
\noindent
FIGURE CAPTIONS

\noindent
Fig.1.    The effective potential is plotted for the dimensionless
 parameter:  \,  $\beta=$ 0.068 (top curve), 0.072,
 and 0.078 (bottom curve); the maximum value is located at 
 $r_{un}=(1-\sqrt{1-12\beta})/2$.\\

\noindent
Fig.2.  A graphic of the positive branch of Eq. (\ref{tau}) 
is pictured  for  values of the
 parameter $\beta=$ 0.064 (top curve), 0.065, 0,066, 0,067,
 and 0.068 (bottom curve).  \\
 
\noindent
Fig.3.  A graphic of \, $\sin [\Omega\tau(r)] $
 for $\beta=0.064$, and $\Omega$ =0.05 (top curve), 0.06, 0.07, and 0.08
 (bottom curve). \\

\noindent Fig.4.   Plot of the integrand of $K(\Omega)$  for $\beta$=0.064,
 $\Omega$=0.06, and
  $f'(r) \equiv df/dr=r$ (top curve), $r^2/10$, and $r^3/100$ (bottom curve).\\

\noindent Fig.5.  Plot of the integrand of $K(\Omega)$  for 
$\beta$=0.068, $\Omega$=0.15, and $f'(r) =1/(3r^2)$ (top curve),
 $1/(9r^3)$, and $1/((27r^4),$ (bottom curve).

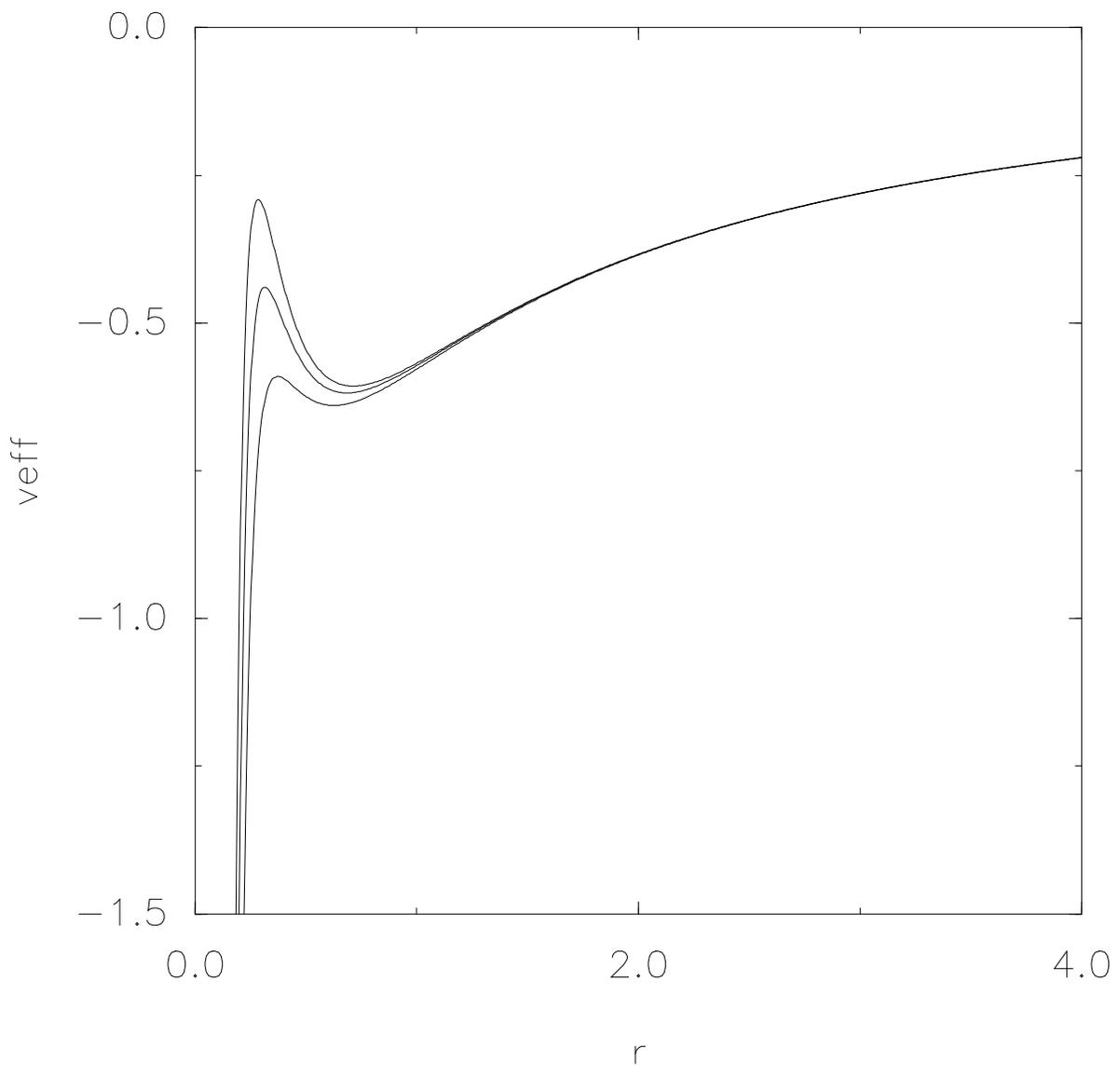
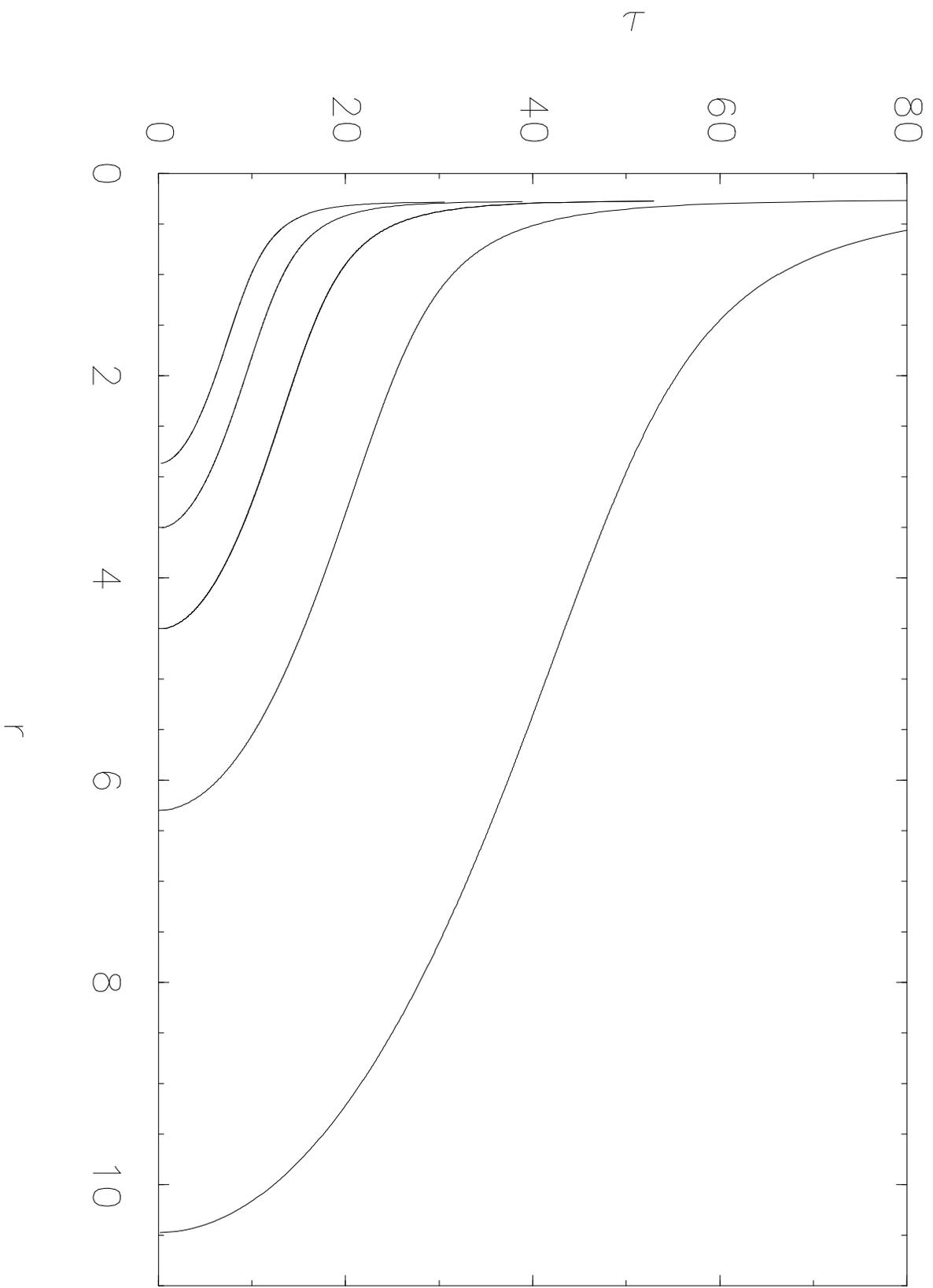
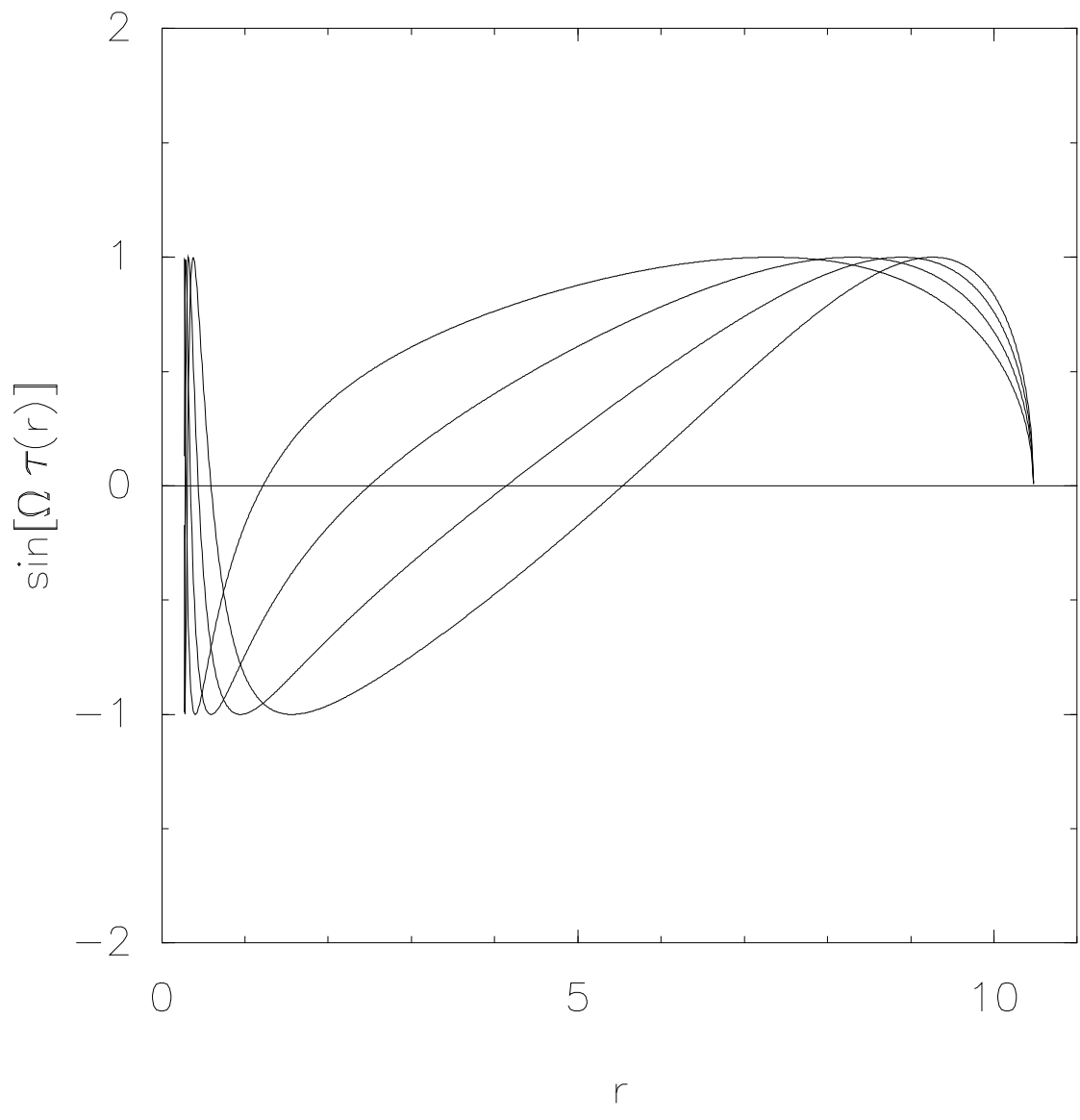
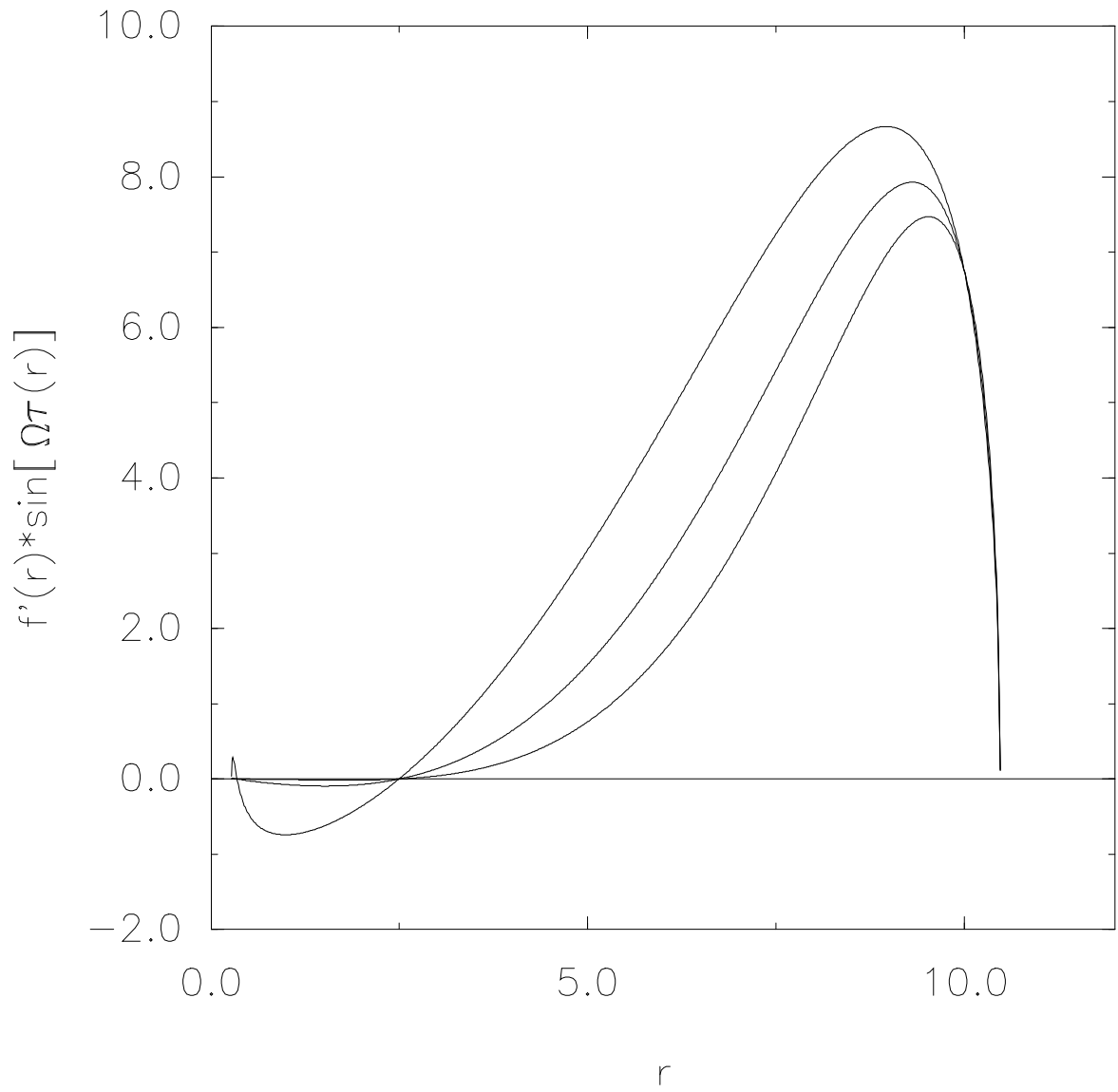
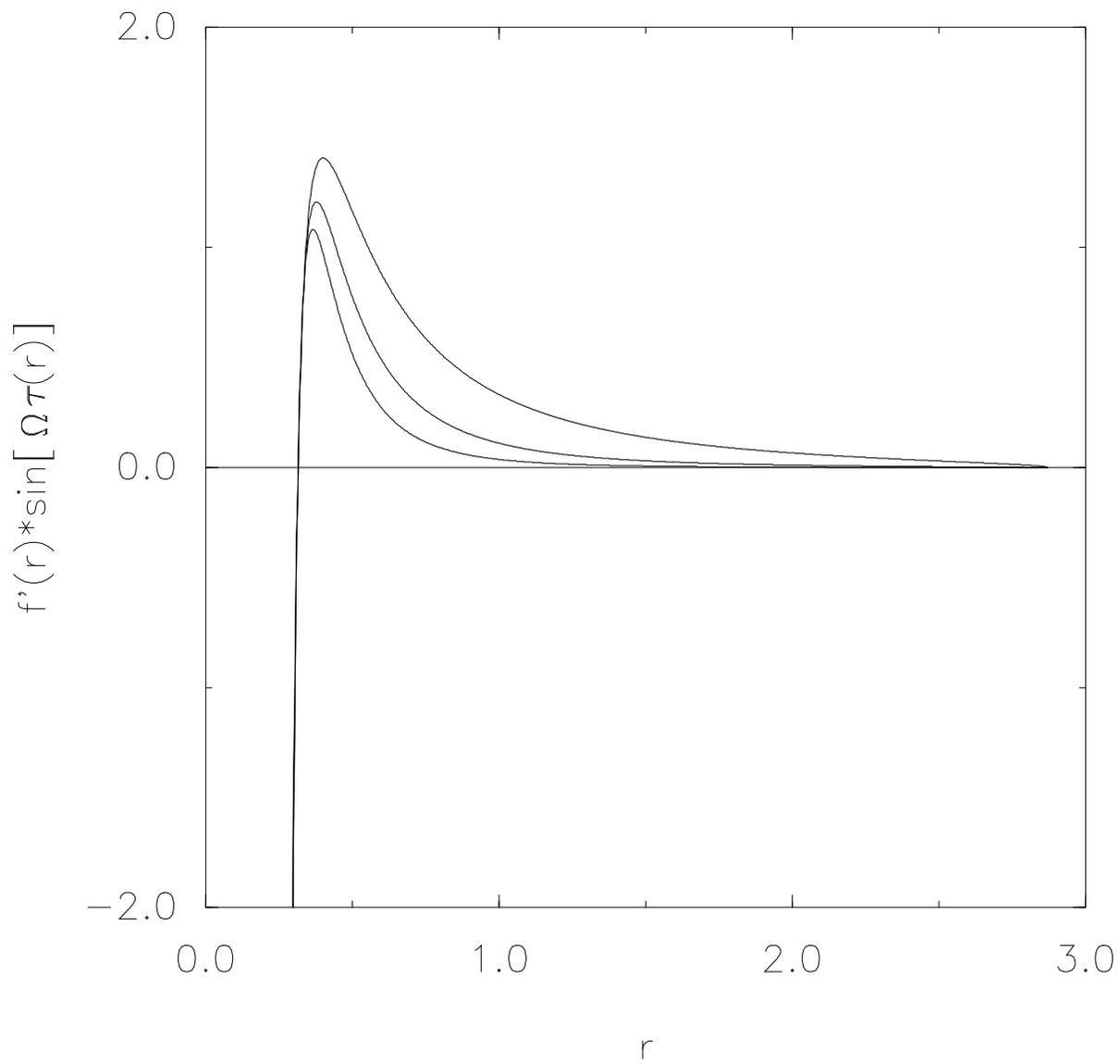
\end{document}